\documentclass[11pt]{JHEP3}
\usepackage{subfigure,axodraw, amsmath,graphicx}

\renewcommand{\vec}[1]{{\mathbf{#1}}}

\newcommand{\be}{\begin{equation}}
\newcommand{\ee}{\end{equation}}
\newcommand{\bq}{\begin{eqnarray}}
\newcommand{\eq}{\end{eqnarray}}

\title{The MHV lagrangian vertices and the Parke-Taylor formula
}

\author{ Chih-Hao Fu\\   
Department of Mathematical Sciences, University of Durham\\
South Road, Durham, DH1 3LE, U.K.\\ 
E-mails:
 \email{chih-hao.fu@durham.ac.uk} } 

\maketitle

\preprint{SHEP 09-08}

\abstract{ 
We explicitly calculate the vertices of the MHV-rules lagrangian in
4-dimensions. This proves that the vertices in the lagrangian obtained by a canonical transformation from light-cone Yang-Mills theory coincide to all order with the Parke-Taylor formula, 
filling the gap originally left in the lagrangian derivation of the CSW rules.
}

\begin{document}

\section{Introduction}

The standard perturbative calculation for pure Yang-Mills theory
is known to be challenging because the number of Feynman diagrams 
contributing to a scattering amplitude 
increases rapidly with the number of legs even at tree-level. A number
of techniques has been devised to simplify the task. In \cite{ParkeTaylor} 
Parke and Taylor conjectured the general formula for colour-ordered
MHV n-gluon scattering amplitudes. (Amplitudes with the most 
plus helicity gluons possible.) The formula was later proved by Berends and
Giele \cite{BerendsGiele} using recursion methods. In \cite{Cachazo:2004kj} Cachazo,
Svr$\check{c}$ek and Witten discovered a set of remarkably simple
rules inspired by twistor string theory to calculate scattering
amplitudes with generic helicity configurations. The CSW rules take
off-shell continued MHV amplitudes and scalar propagators as their building
blocks and have been successfully applied at tree level \cite{Georgiou:2004wu, Bena:2004ry, 
Birthwright:2005ak} and several loop-level amplitudes \cite{Bidder:2005ri, Roiban:2004ix, Bedford:2004py, Brandhuber:2007vm}.
The proof of the CSW rules however, was provided using another approach.
Britto, Cachazo, Feng and Witten \cite{Britto:2004ap, Britto:2005fq} derived a new recursion
relation by analysing singularities of scattering amplitudes.
Using Cauchy's theorem, it was shown that the analytically continued
amplitude can be equally obtained from a sum over residues \cite{Luo:2005my, Britto:2004nc, Bern:2005hs, Bertolini:2005di, Bern:2005hh}.
The BCFW recursion method has been generalised to include massive
particles \cite{Badger:2005zh, Badger:2005jv, Forde:2005ue, Ozeren:2006ft}, superpartners \cite{Bianchi:2008pu} and to
 gravity \cite{Bedford:2005yy, Cachazo:2005ca}.

In \cite{Mansfield:2005yd} a lagrangian derivation of CSW rules was found. Starting
from the Yang-Mills lagrangian in light-cone gauge, a canonical
transformation was applied to transverse components of the gauge
field to rearrange the self-dual part of the lagrangian into a free
field theory

\[
\mathcal{L}^{-+}\left[\mathcal{A}\right]+\mathcal{L}^{--+}\left[\mathcal{A}\right]=\mathcal{L}^{-+}\left[\mathcal{B}\right]\]

After the transformation the equivalent lagrangian theory contains only
vertices that have maximum helicity contents, in agreement with the
CSW prescription. The method was extended to QCD
\cite{Ettle:2008ey} and supersymmetric theories \cite{Feng:2006yy, Morris:2008uc}
In particular, a corresponding D-dimensional MHV-rules lagrangian
has been developed to incorporate dimensional regularisation and to
explain the non-vanishing all-plus amplitudes at one-loop level which
do not appear in the CSW construction \cite{Ettle:2007qc, Fu:2009nh}. Alternatively,
one can choose to work in 4-dimension provided a suitable regulator
is imposed. In \cite{Brandhuber:2007vm} Brandhuber, Spence, Travaglini and Zoubos used the 
light-cone friendly regularisation scheme of Thorn \cite{Thorn:2005ak}. 
In this approach the ``missing'' all-plus
amplitudes were explained by the extra contribution from counterterms.

In \cite{Mansfield:2005yd} an indirect argument was given to show that in
 4-dimension, the vertices of the MHV-rules lagrangian were argued
to have the same form as the Parke-Taylor formula.
At tree level, scattering amplitudes and the vertices can only differ
by factors that contain squares of momenta, which vanish on-shell.
Since the vertices in 4-dimension were known to be holomorphic such
factors must be absent\footnote{
Note that the holomorphic behaviour of the MHV vertices is exclusive to
 the choice implicitly taken by the canonical transformation,
where the canonical conjugate momentum $\hat{p}\bar{\mathcal{A}}(p)$
is assumed to have the inverse transformation relation from the field
variable $\mathcal{A}(p)$. One can show that for a generic function
of momentum $f(p)$, the transformation 

\begin{equation}
f(p)\bar{\mathcal{A}}^{a}(p)=\int d^{4}q\,\frac{\delta\mathcal{B}^{b}(q)}{\delta\mathcal{A}^{a}(p)}\, f(q)\bar{\mathcal{B}}(q)
\label{general transform}
\end{equation}

also preserves the integration measure. Following the method originally
used in \cite{Mansfield:2005yd} one can obtain vertices that give the
same helicity configurations as the vertices used in the CSW rules.
The generic translation kernel derived from the condition (\ref{general transform})
is not holomorphic, and cannot be expressed as products of round brackets.
The MHV vertices derived from general measure-preserving transformations
are not the same as the Parke-Taylor formula. 
}. Explicit calculations verified that the
vertices agree with the Parke-Taylor formula up to 5-points \cite{Ettle:2006bw}, however a general verification
to all vertices has been missing. In this short paper we present the
proof to show that n-point vertices match with the Parke-Taylor formula.
The notation used throughout the paper follow the definitions in
\cite{Ettle:2006bw} and is summarised in the appendix.

\section{MHV vertices in 4 dimensions and the Parke-Taylor formula}

In the MHV lagrangian theory \cite{Mansfield:2005yd} components of the gauge field
$\mathcal{A}$ and $\bar{\mathcal{A}}$ in the light-cone coordinates
are expanded into a new set of field variables $\mathcal{B}$ and $\bar{\mathcal{B}}$
using a canonical transformation. The new vertices are then derived
by translating the $\mathcal{A}$ and $\bar{\mathcal{A}}$ fields
attached to the vertex terms in the original lagrangian into $\mathcal{B}$
and $\bar{\mathcal{B}}$. For the 4-dimensional theory, Ettle and
Morris \cite{Ettle:2006bw} have shown that the n-th order term in the $\mathcal{A}$
field expansion can be summarised by a simple formula.


\begin{figure}[!h]
  \centering \subfigure{
  \begin{picture}(108,64) (9,7)
    \SetWidth{0.375}

    \Line(15,34)(36,34)

    \Line(45,34)(72,13)
    \Line(45,34)(75,52)
    \Text(12,31)[br]{$\mathcal{A}_{1} $}
    
    \Text(93,54)[br]{$ \mathcal{B}_{2} $}
    \Text(85,37)[br]{.}
    \Text(85,34)[br]{.}
    \Text(85,31)[br]{.}
    \Text(85,28)[br]{.}
    \Text(93,3)[br]{$ \mathcal{B}_{n}$}
       \BCirc(45,34){8}
  \end{picture}}

\end{figure}

\begin{equation}
\mathcal{A}_{1}\rightarrow\frac{\hat{1}\,\hat{3}\hat{4}\cdots\widehat{n-1}}{\left(23\right)\cdots\left(n-1,n\right)}\,\mathcal{B}_{2}\mathcal{B}_{3}\cdots\mathcal{B}_{n}\end{equation}

Similarly, the $\bar{\mathcal{A}}$ expansion was shown to have the
form \cite{Fu:2009nh}


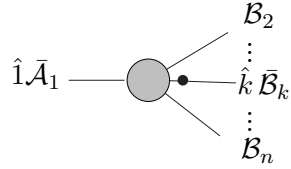
\begin{figure}[!h]
  \centering \subfigure{
  \begin{picture}(108,64) (9,7)
    \SetWidth{0.375}

    \Line(15,34)(36,34)

    \Line(45,34)(72,13)
    \Line(45,34)(75,52)
    \Line(48,34)(78,33)
    \Text(12,31)[br]{$\hat{1}\bar{\mathcal{A}}_{1} $}
    \Text(85,44)[br]{.}
    \Text(85,47)[br]{.}
    \Text(85,41)[br]{.}
    
    \Text(93,54)[br]{$ \mathcal{B}_{2} $}
    
    \Text(85,21)[br]{.}
    \Text(85,18)[br]{.}
    \Text(85,15)[br]{.}
    \Text(99,27)[br]{$ \hat{k}\, \bar{\mathcal{B}}_{k}$}
    \Text(93,3)[br]{$ \mathcal{B}_{n}$}
       \GCirc(45,34){8}{0.75}
           \Vertex(58,34){2}
  \end{picture}}

\end{figure}

\begin{equation}
\hat{1}\bar{\mathcal{A}}_{1}\rightarrow\frac{\hat{k}\,\hat{3}\hat{4}\cdots\widehat{n-1}}{\left(23\right)\cdots\left(n-1,n\right)}\,\mathcal{B}_{2}\mathcal{B}_{3}\cdots\left(\hat{k}\bar{\mathcal{B}}_{k}\right)\cdots\mathcal{B}_{n}\end{equation}

In the following we shall prove that the new vertices have the same form as
the Parke-Taylor formula \cite{ParkeTaylor} by first proving that the MHV vertices
described above and the Parke-Taylor formula can both be spanned by
terms of the form

\begin{eqnarray}
&& \frac{1}{\left(23\right)\left(34\right)\cdots\left(k-1,k\right)}\times\frac{1}{\left(k+1,k+2\right)\cdots\left(m-1,m\right)} \nonumber\\
&& \times\frac{1}{\left(m+1,m+2\right)\cdots\left(l-1,l\right)}\times\frac{1}{\left(l+1,l+2\right)\cdots\left(n,1\right)}
\label{eq1}
\end{eqnarray}

together with terms of the form

\begin{equation}
\frac{\left(12\right)}{\left(23\right)\left(34\right)\cdots\left(k-1,k\right)}\times\frac{1}{\left(k+1,k+2\right)\cdots\left(l-1,l\right)}\times\frac{1}{\left(l+1,l+2\right)\cdots\left(n,1\right)}
\label{eq2}
\end{equation}

and then we shall check that the coefficients of the expansion agree with each
other. The denominators of (\ref{eq1}) and (\ref{eq2}) contain only
round brackets of adjacent legs and are split into three and four
groups of sequential products respectively. (Note however, the product
is taken as $1$ if it starts and ends at the same leg.) The vertices
and the Parke-Taylor formula are regarded as functions of tilde component
variables $\tilde{p}$ contained in the round brackets while expansion
coefficients depend only on hat components $\hat{p}$. For example,
the 5-point Parke-Taylor formula can be rewritten as

\begin{eqnarray}
&& \frac{\left(12\right)^{3}}{\left(23\right)\left(34\right)\left(45\right)\left(51\right)}=A\,\frac{\left(12\right)}{\left(23\right)\left(34\right)}+B\,\frac{\left(12\right)}{\left(23\right)\left(45\right)}+C\,\frac{\left(12\right)}{\left(23\right)\left(51\right)}   \nonumber \\
&& \hspace{3.2cm}  +D\,\frac{\left(12\right)}{\left(34\right)\left(45\right)}+E\,\frac{\left(12\right)}{\left(34\right)\left(51\right)}+F\,\frac{\left(12\right)}{\left(45\right)\left(51\right)}   \nonumber\\
&& \hspace{3.2cm} +G\,\frac{1}{\left(23\right)}+H\,\frac{1}{\left(34\right)}+I\,\frac{1}{\left(45\right)}+J\,\frac{1}{\left(51\right)}\end{eqnarray}

The coefficients can be easily determined by the method of
 partial fractions. To calculate $A$ we set $\left(23\right)$
and $\left(34\right)$ to be zero. These conditions allow us to solve
$\tilde{3}$ and $\tilde{4}$ in terms of $\tilde{2}$.

\begin{equation}
\tilde{3}=\frac{\hat{3}}{\hat{2}}\tilde{2}, \hspace{2cm}  \tilde{4}=\frac{\hat{4}}{\hat{2}}\tilde{2}\end{equation}

Brackets formed by momenta $3$ and $4$ with other legs $p$ $q$
can therefore be replaced by brackets of $2$ with $p$ $q$.

\begin{equation}
\left(3,\, p\right)=\frac{\hat{3}}{\hat{2}}\left(2,\, p\right),\hspace{0.5cm}  \left(4,\, q\right)=\frac{\hat{4}}{\hat{2}}\left(2,\, q\right)\end{equation}

Together with momentum conservation the remaining brackets $\left(45\right)$
and $\left(51\right)$ can be expressed in terms of $\left(12\right)$.
Matching both sides of the equation gives us the coefficient $A$. For
terms like $G$ that do not have $\left(12\right)$ in the numerator
we set $\left(12\right)$ and $\left(23\right)$ zero. The other coefficients
are then determined through the same procedure.

\begin{eqnarray}
A=\frac{\hat{2}\hat{3}\hat{5}}{\hat{1}\left(\hat{2}+\hat{3}+\hat{4}\right)},\, B=\frac{\hat{2}\left(\hat{1}+\hat{2}+\hat{3}\right)^{2}}{\hat{1}\left(\hat{2}+\hat{3}\right)},\, C=\frac{\hat{1}\hat{2}\hat{4}}{\left(\hat{1}+\hat{5}\right)\left(\hat{2}+\hat{3}\right)}, \\
D=\frac{\hat{4}\left(\hat{3}+\hat{4}+\hat{5}\right)^{2}}{\hat{1}\hat{2}},\, E=\frac{-\hat{1}\hat{4}\left(\hat{1}+\hat{2}+\hat{5}\right)^{2}}{\hat{2}\hat{3}\left(\hat{1}+\hat{5}\right)},\, F=\frac{\hat{1}\hat{3}\hat{5}}{\hat{2}\left(\hat{1}+\hat{4}+\hat{5}\right)}\end{eqnarray}

\begin{equation}
G=H=I=J=0\end{equation}

From the method described above, it is clear that the Parke-Taylor
formula does not contribute to terms independent of $\left(12\right)$.
At the end of the argument we shall show this is generally also true
for the n-point MHV vertices, but for convenience for the moment we
will retain such terms in the expansion.

\subsection{Partial fraction expansion}

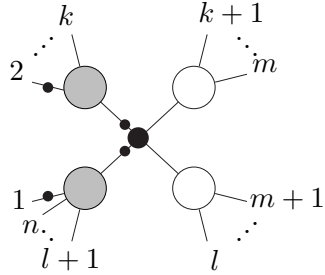
\begin{figure}[!h]
  \centering 
  \subfigure{
    \begin{picture}(0,5) 
    \end{picture}
  }
  \\
    \subfigure{

  \begin{picture}(81,78) (24,-9)
    \SetWidth{0.375}

    \Line(45,48)(84,12)
    \Line(45,12)(84,48)

    \Line(87,57)(90,69)
    \Line(93,51)(105,54)
    \Line(24,6)(36,9)
    \Line(39,-9)(42,3)
    \Line(42,57)(39,69)
    \Line(36,48)(24,51)
    \Line(90,-9)(87,3)
    \Line(105,6)(93,9)
    \SetWidth{0.5}
    \Vertex(59,25){2}
    \Vertex(59,35){2}
    \Vertex(64,30){4}
    \SetWidth{0.375}
    \Line(28,1)(37,7)

    \GCirc(44,49){8}{0.75}
    \GCirc(44,11){8}{0.75}    
    \BCirc(85,11){8}
    \BCirc(85,49){8}
    
        \Text(22,3)[br]{$1$}
        \Text(27,-6)[br]{$n$}
        \Text(50,-20)[br]{$l+1$}
        \Text(30,-6)[br]{.}
        \Text(33,-9)[br]{.}
                \Text(95,-20)[br]{$l$}
                \Text(135,3)[br]{$m+1$}
                \Text(110,-3)[br]{.}
                \Text(107,-6)[br]{.}
                \Text(104,-9)[br]{.}
                          \Text(112,73)[br]{$k+1$}
                          \Text(117,55)[br]{$m$}
                          \Text(110,61)[br]{.}
                          \Text(107,64)[br]{.}
                          \Text(104,67)[br]{.}
                                \Text(40,73)[br]{$k$}
                                \Text(21,52)[br]{$2$}
                                \Text(27,61)[br]{.}
                                \Text(30,64)[br]{.}
                                \Text(33,67)[br]{.}
        \Vertex(30,49){2}
         \Vertex(30,8){2}
  \end{picture} 
    
  }   \caption{Translated 4-point vertex}
\label{spider graph}
  \end{figure}

\begin{figure}[!h]
\centering
  \subfigure{
    \begin{picture}(0,0) (10,10)
    \end{picture}
  }
  \\
  \subfigure{
    \begin{picture}(81,78) (24,-9)
    \SetWidth{0.375}

    \Line(45,48)(64,30)
        \Line(64,30)(85,30)
    \Line(45,12)(64,30)

    \Line(90,34)(102,42)
    \Line(24,6)(36,9)
    \Line(39,-9)(42,3)
    \Line(42,57)(39,69)
    \Line(36,48)(24,51)

    \Line(102,18)(90,26)
    \SetWidth{0.5}
    \Vertex(59,25){2}
    \Vertex(59,35){2}
    \Vertex(64,30){4}
    \SetWidth{0.375}
    \Line(28,1)(37,7)

    \GCirc(44,49){8}{0.75}
    \GCirc(44,11){8}{0.75}    
    \BCirc(85,30){8}
    
        \Text(22,3)[br]{$1$}
        \Text(27,-6)[br]{$n$}
        \Text(50,-20)[br]{$l+1$}
        \Text(30,-6)[br]{.}
        \Text(33,-9)[br]{.}
                \Text(125,47)[br]{$k+1$}
                \Text(110,13)[br]{$l$}
                \Text(110,33)[br]{.}
                \Text(110,30)[br]{.}
                \Text(110,27)[br]{.}

                                \Text(40,73)[br]{$k$}
                                \Text(21,52)[br]{$2$}
                                \Text(27,61)[br]{.}
                                \Text(30,64)[br]{.}
                                \Text(33,67)[br]{.}
        \Vertex(30,49){2}
        \Vertex(30,8){2}
  \end{picture}
  
  }  \caption{Translated 3-point vertex} \label{lobster graph}
\end{figure}
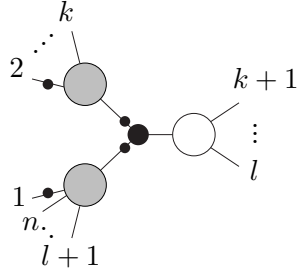

To justify the expansion we need to show (\ref{eq1}) and (\ref{eq2})
are sufficient to describe MHV vertices and the Parke-Taylor formula.
An n-point vertex in the MHV lagrangian theory consists of terms splitted
from the 3-point and 4-point LCYM vertices. The contributions from
the 4-point vertex (Fig.\ref{spider graph}) naturally are of the form (\ref{eq1}).
For vertices that originate from the 3-point vertex (Fig.\ref{lobster graph}), translating
$\mathcal{A}$ and $\bar{\mathcal{A}}$ fields associated with each
leg into $\mathcal{B}$ and $\bar{\mathcal{B}}$ produces a series
of products of brackets. Using the bilinear property the factor $\left(1+\cdots\left(l+1\right),\,2+\cdots k\right)$ in the numerator can be expanded into brackets of single leg momenta
$\left(p,q\right)$ with $p$ and $q$ running through $1$ to $l+1$
and $2$ to $k$ respectively. Each term $\left(p,q\right)$ can then
be rewritten as a linear combination of brackets of adjacent momenta
by noticing that

\begin{eqnarray}
\frac{(p,q)}{\hat{p}\hat{q}}=\frac{\tilde{q}}{\hat{q}}-\frac{\tilde{p}}{\hat{p}}=\frac{\tilde{q}}{\hat{q}}-\frac{\widetilde{p-1}}{\widehat{p-1}}+\frac{\widetilde{p-1}}{\widehat{p-1}}-\frac{\tilde{p}}{\hat{p}} \nonumber \\ 
=\frac{\left(p-1,q\right)}{\widehat{p-1}\hat{q}}+\frac{\left(p,p-1\right)}{\hat{p}\widehat{p-1}}
\hspace{1.5cm}
\label{eq3}
\end{eqnarray}

Applying (\ref{eq3}) repeatedly $p$ and $q$ can be moved toward
$1$ and $2$, resulting in a term of the form (\ref{eq2}) while
brackets of adjacent momenta produced in the process cancel brackets
in the denominator, resulting terms of the form (\ref{eq1}).


The Parke-Taylor formula can also be spanned by (\ref{eq1}) and (\ref{eq2}).
To show this is true we need to express two of the $\left(12\right)$
factors in the numerator as a linear combination of products of two
different brackets $\left(ab\right)\left(cd\right)$. To replace a
first $\left(12\right)$, notice that the momentum labels are defined cyclically

\begin{equation}
\sum_{k=1}^{n}\frac{\left(k,k+1\right)}{\hat{k}\widehat{k+1}}=\sum_{k=1}^{n}\left(\frac{\widetilde{k+1}}{\widehat{k+1}}-\frac{\tilde{k}}{\hat{k}}\right)=0\end{equation}

Therefore we have

\begin{equation}
\frac{\left(12\right)}{\hat{1}\hat{2}}=-\frac{\left(23\right)}{\hat{2}\hat{3}}-\frac{\left(34\right)}{\hat{3}\hat{4}}-\cdots\frac{\left(n1\right)}{\hat{n}\hat{1}}
\label{eq4}
\end{equation}

Using (\ref{eq4}) to expand one of the $\left(12\right)$ factors
gives us a sum over terms of the form $\left(12\right)^{2}\left(ab\right)$.
Another $\left(12\right)$ can be replaced by first applying conservation
of momentum to substitute one of the legs

\begin{equation}
\left(12\right)=-\left(13\right)-\left(14\right)-\cdots\left(1n\right)
\label{eq5}
\end{equation}

Applying (\ref{eq3}) again, all of the round brackets on the right
hand side can be expressed in terms of brackets of adjacent momenta.
We then have two equations (\ref{eq4}) and (\ref{eq5}) for $\left(12\right)$
and $\left(ab\right)$. Solving the equations will give us an expression
for $\left(12\right)$ in terms of brackets other than $\left(ab\right)$,
which can be used to replace the second $\left(12\right)$ in the
numerator of the Parke-Taylor formula.

\subsection{Matching expansion coefficients}

Since the MHV vertices and the Parke-Taylor formula are spanned by
functions of round brackets with coefficients depending on hat components
only, as in the 5-point case shown at the beginning of this section
we are free to adjust all of the tilde component variables on both sides
of the expansion equation to solve for the coefficients. First let
us check the coefficients of (\ref{eq2}). For the n-point MHV vertex,
the contributions to (\ref{eq2}) come solely from terms translated
from the 3-point LCYM vertex (Fig. \ref{lobster graph}). Following the convention introduced
in \cite{Fu:2009nh} for graphical notation this gives

\begin{eqnarray}
&&\frac{\hat{2}}{\hat{2}+\cdots\hat{k}}\,\frac{\hat{2}\,\hat{3}\cdots\widehat{k-1}}{\left(23\right)\cdots\left(k-1,k\right)}\,\frac{\hat{1}}{\widehat{l+1}+\cdots\hat{1}}\,\frac{\widehat{k+2}\cdots\widehat{l-1}}{\left(l+1,l+2\right)\cdots\left(n1\right)}\nonumber\\
&& \times \frac{\hat{1}\,\widehat{l+2}\cdots\hat{n}}{\left(k+1,k+2\right)\cdots\left(l-1,l\right)} \nonumber \\
&& \times\frac{\widehat{k+1}+\cdots\hat{l}}{\left(\widehat{l+1}+\cdots\hat{1}\right)\left(\hat{2}+\cdots\hat{k}\right)}\left((l+1)+\cdots1,\,2+\cdots k\right)
\label{eq6}
\end{eqnarray}

Using conditions

\begin{equation}
\tilde{k}=\frac{\hat{k}}{\widehat{k-1}}\widetilde{k-1},\hspace{0.5cm} \cdots \hspace{0.5cm} \tilde{3}=\frac{\hat{3}}{\hat{2}}\tilde{2}\end{equation}

\begin{equation}
\widetilde{l+1}=\frac{\widehat{l+1}}{\widehat{l+2}}\widetilde{l+2}, \hspace{0.4cm} \cdots \hspace{0.4cm} \tilde{n}=\frac{\hat{n}}{\hat{1}}\tilde{1}\end{equation}

\begin{equation}
\hspace{0.5cm} \widetilde{k+1}=\frac{\widehat{k+1}}{\widehat{k+2}}\widetilde{k+2}, \hspace{0.3cm} \cdots \hspace{0.3cm} \widetilde{l-1}=\frac{\widehat{l-1}}{\hat{l}}\tilde{l}\end{equation}

and conservation of momentum

\begin{equation}
\tilde{l}=-\tilde{1}-\tilde{2}-\cdots-\widetilde{l-1}-\widetilde{l+1}\cdots-\tilde{n}\end{equation}

the numerator simplifies to

\begin{equation}
\left((l+1)+\cdots1,\,2+\cdots k\right)=\frac{\left(\widehat{l+1}+\cdots\hat{1}\right)\left(\hat{2}+\cdots\hat{k}\right)}{\hat{1}\hat{2}}\left(12\right)\end{equation}

Similarly, for the expanded Parke-Taylor formula we have 

\begin{equation}
\left(k,k+1\right)=\frac{\hat{k}}{\hat{2}}\,\frac{\widehat{k+1}}{\hat{l}}\,\frac{\left(\widehat{l+1}+\cdots\hat{1}\right)}{\hat{1}}\,\frac{\hat{l}}{\left(\widehat{k+1}+\cdots\hat{l}\right)}\left(12\right)\end{equation}

\begin{equation}
\left(l,l+1\right)=\frac{\widehat{l+1}}{\hat{1}}\,\frac{\left(\hat{k}+\cdots\hat{2}\right)}{\hat{2}}\,\frac{\hat{l}}{\left(\widehat{k+1}+\cdots\hat{l}\right)}\left(12\right)\end{equation}

Collecting terms, both (\ref{eq6}) and the Parke-Taylor formula give
the same coefficient for (\ref{eq2}).

\begin{equation}
\frac{\hat{1}\hat{2}\cdots\hat{n}}{\hat{k}\widehat{k+1}\,\hat{l}\,\widehat{l+1}}\,\frac{\left(\widehat{k+1}+\cdots\hat{l}\right)^{2}}{\left(\hat{k}+\cdots\hat{2}\right)\left(\widehat{l+1}+\cdots\hat{1}\right)}\end{equation}

\begin{figure}[!h]
  \centering 
  \subfigure{
    \begin{picture}(5,5) 
    \end{picture}
  }
  \\
  \subfigure{
    \begin{picture}(5,78) 
                \Text(0,75)[br]{$(a)$}
    \end{picture}
  }
    \subfigure{
      \begin{picture}(81,78) (24,-9)
    \SetWidth{0.375}

    \Line(45,48)(64,30)
        \Line(64,30)(85,30)
    \Line(45,12)(64,30)

    \Line(90,34)(102,42)
    \Line(24,6)(36,9)
    \Line(39,-9)(42,3)
    \Line(42,57)(39,69)
    \Line(36,48)(24,51)


    \Line(102,18)(90,26)
    \SetWidth{0.5}
    \Vertex(59,25){2}
    \Vertex(59,35){2}
    \Vertex(64,30){4}
    \SetWidth{0.375}
    \Line(28,1)(37,7)
    \Line(34,-5)(39,5)    
    
    \GCirc(44,49){8}{0.75}
    \GCirc(44,11){8}{0.75}    
    \BCirc(85,30){8}
    
        \Text(22,3)[br]{$1$}
        \Text(23,-9)[br]{$l+1$}
        \Text(32,-13)[br]{$l$}
        \Text(65,-20)[br]{$m+1$}
                \Text(125,47)[br]{$k+1$}
                \Text(120,13)[br]{$m$}
                \Text(110,33)[br]{.}
                \Text(110,30)[br]{.}
                \Text(110,27)[br]{.}

                                \Text(40,73)[br]{$k$}
                                \Text(21,52)[br]{$2$}
                                \Text(27,61)[br]{.}
                                \Text(30,64)[br]{.}
                                \Text(33,67)[br]{.}
           \Vertex(30,49){2}
            \Vertex(30,8){2}
  \end{picture}    
}
  \subfigure{
    \begin{picture}(40,0) 
    \end{picture}
  }
  \subfigure{
    \begin{picture}(10,78) 
                \Text(0,75)[br]{$(b)$}
    \end{picture}
  }   
 \subfigure{
      \begin{picture}(81,78) (24,-9)
    \SetWidth{0.375}

    \Line(45,48)(64,30)
        \Line(64,30)(85,30)
    \Line(45,12)(64,30)

    \Line(90,34)(102,42)
    \Line(24,6)(36,9)
    \Line(39,-9)(42,3)
    \Line(42,57)(39,69)
    \Line(36,48)(24,51)
        \Line(29,57)(38,51)
        \Line(34,63)(40,54)

    \Line(102,18)(90,26)
    \SetWidth{0.5}
    \Vertex(59,25){2}
    \Vertex(59,35){2}
    \Vertex(64,30){4}
    \SetWidth{0.375}

    \GCirc(44,49){8}{0.75}
    \GCirc(44,11){8}{0.75}    
    \BCirc(85,30){8}
    
        \Text(22,3)[br]{$1$}
        \Text(50,-20)[br]{$l+1$}
        \Text(27,-3)[br]{.}
        \Text(30,-6)[br]{.}
        \Text(33,-9)[br]{.}
                \Text(125,47)[br]{$m+1$}
                \Text(110,13)[br]{$l$}
                \Text(110,33)[br]{.}
                \Text(110,30)[br]{.}
                \Text(110,27)[br]{.}

                                \Text(45,73)[br]{$m$}
                                \Text(18,46)[br]{$2$}
                                    \Text(26,57)[br]{$k$}
                                    \Text(37,66)[br]{$k+1$}
          \Vertex(30,49){2}
           \Vertex(30,8){2}
  \end{picture}    
}
\\
  \subfigure{
    \begin{picture}(5,10) 
    \end{picture}
  }
  \\
  \subfigure{
    \begin{picture}(10,78) 
        \Text(0,75)[br]{$(c)$}
    \end{picture}
  } 
   \subfigure{
      \begin{picture}(81,78) (24,-9)
    \SetWidth{0.375}

    \Line(45,48)(64,30)
        \Line(64,30)(85,30)
    \Line(45,12)(64,30)

    \Line(90,34)(102,42)
    \Line(24,6)(36,9)
    \Line(39,-9)(42,3)
    \Line(42,57)(39,69)
    \Line(36,48)(24,51)

        \Line(90,29)(102,26)
        \Line(90,31)(102,34)

    \Line(102,18)(90,26)
    \SetWidth{0.5}
    \Vertex(59,25){2}
    \Vertex(59,35){2}
    \Vertex(64,30){4}
    \SetWidth{0.375}
    
    \GCirc(44,49){8}{0.75}
    \GCirc(44,11){8}{0.75}    
    \BCirc(85,30){8}
    
        \Text(22,3)[br]{$1$}
                   \Text(27,-3)[br]{.}
                   \Text(30,-6)[br]{.}
                   \Text(33,-9)[br]{.}
                   
        \Text(55,-20)[br]{$l+1$}
                \Text(125,47)[br]{$k+1$}
                    \Text(133,23)[br]{$m+1$}
                    \Text(116,35)[br]{$m$}
                \Text(110,11)[br]{$l$}

                                \Text(40,73)[br]{$k$}
                                \Text(21,52)[br]{$2$}
                                \Text(27,61)[br]{.}
                                \Text(30,64)[br]{.}
                                \Text(33,67)[br]{.}
         \Vertex(30,49){2}
          \Vertex(30,8){2}
  \end{picture}    
}
  \label{lobsters abc}
  \caption{Contributions from the 3-point vertex}
\end{figure}
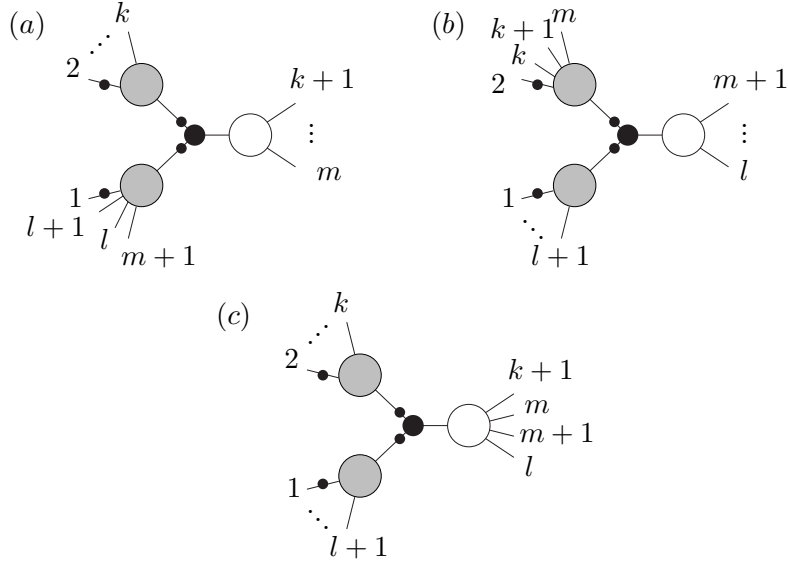

As for the coefficient of terms (\ref{eq1}), we receive contributions
from graphs translated from the 4-point vertex (Fig.\ref{spider graph}), for which
the translation kernels from the legs yield factors of the form (\ref{eq1}),
and contributions from graphs using the 3-point vertex as backbone
(Fig.\ref{lobsters abc} (a) to (c)), in which case the bracket in the numerator cancels
another bracket coming from the kernel and splits the denominator
into two sets of sequential products of brackets.

For simplicity we extract the common factors from each graph

\begin{eqnarray}
&& \frac{\hat{2}\,\hat{3}\cdots\widehat{k-1}}{\left(23\right)\cdots\left(k-1,k\right)}\times\frac{\widehat{k+2}\cdots\widehat{m-1}}{\left(k+1,k+2\right)\cdots\left(m-1,m\right)} \nonumber \\
&& \times\frac{\widehat{m+2}\cdots\widehat{l-1}}{\left(m+1,m+2\right)\cdots\left(l-1,l\right)}\times\frac{\hat{1}\,\widehat{l+2}\cdots\hat{n}}{\left(l+1,l+2\right)\cdots\left(n,1\right)}
\end{eqnarray}

The remaining factors are then simplified by partial fractions. For
graph (a), this is

\begin{equation}
\frac{\hat{l}\,\widehat{l+1}}{\left(l,l+1\right)}\,\frac{\left((m+1)+\cdots1,\,2+\cdots k\right)}{\left(\widehat{m+1}+\cdots\hat{1}\right)\left(\hat{2}+\cdots\hat{k}\right)}\,\frac{\hat{1}\hat{2}\left(\widehat{k+1}+\cdots\hat{m}\right)}{\left(\widehat{m+1}+\cdots\hat{1}\right)\left(\hat{2}+\cdots\hat{k}\right)}=\frac{\hat{1}\hat{2}}{(a+d)^{2}}\,\frac{c^{2}d}{b}
\label{eq7}
\end{equation}

where $a, b, c$ and $d$ denote the momenta of the four lines streching out of the 4-point vertex in (Fig.\ref{spider graph})

\begin{eqnarray}
&& a = \widehat{l+1}+ \cdots +\hat{n} +\hat{1} \\
&& b =  \hat{2} + \hat{3} + \cdots +\hat{k} \\
&& c =  \widehat{k+1}+ \cdots + \hat{m}  \\
&& d = \widehat{m+1} + \cdots +\hat{l}
\end{eqnarray}

Similarly for graph (b) we have

\begin{equation}
\frac{\hat{k}\,\widehat{k+1}}{\left(k,k+1\right)}\,\frac{\left((l+1)+\cdots1,\,2+\cdots m\right)}{\left(\widehat{l+1}+\cdots\hat{1}\right)\left(\hat{2}+\cdots\hat{m}\right)}\,\frac{\hat{1}\hat{2}\left(\widehat{m+1}+\cdots\hat{l}\right)}{\left(\widehat{l+1}+\cdots\hat{1}\right)\left(\hat{2}+\cdots\hat{m}\right)}=\frac{\hat{1}\hat{2}}{(a+d)^{2}}\,\frac{cd^{2}}{a}
\label{eq8}
\end{equation}

After simplification graph (c) is proportional to $\left(12\right)$, and
therefore vanishes

\begin{equation}
\left((l+1)+\cdots1,\,2+\cdots k\right)=\frac{\widehat{l+1}+\cdots\hat{1}}{\hat{1}}\,\frac{\hat{2}+\cdots\hat{k}}{\hat{2}}\left(12\right)=0\end{equation}

Putting (\ref{eq7}) and (\ref{eq8}) together cancels the contribution
from (Fig.\ref{spider graph}) 

\begin{equation}
-\frac{\hat{1}\hat{2}}{(a+d)^{2}}\,\frac{cd}{ab}\,(ac+bd)\end{equation}

we thus verified that all of the expansion coefficients for terms
of the form (\ref{eq1}) are zero, as claimed at the beginning of
the section. Since the argument presented here does not depend on whether the negative helicity legs are adjacent to each other, the result generalises to all n-point vertices.


\section{Conclusion and discussions}

We explicitly proved that a generic n-point vertex of the MHV lagrangian is described by the Parke-Taylor formula, which was originally argued by holomorphism and verified only up to 5-points.  The derivation   presented in this paper also directly showed that the Parke-Taylor formula  defined  by light-cone coordinate external leg momenta through (\ref{spinors}) and (\ref{spinor product}) serves as the off-shell continued MHV amplitude used in the CSW rules derived from the light-cone Yang-Mills lagrangian.

The method described in this paper can also be extended to supersymmetric
theories. Since the translation kernels in QCD and the $N=4$ SYM theory 
\cite{Morris:2008uc, Feng:2006yy} were shown to contain the same 
translation kernel used in the pure YM multiplied by light-cone 
hat components of momenta only, it is straightforward
to apply similar expansions and the method of partial fractions to
verify that the vertices are given by the Parke-Taylor forumla multiplied
by suitable ratios of spinor brackets restricted by SUSY Ward identity.


\appendix
\section{Notation}

In this paper we adopted the shorthand notation $\left(\check{p},\,\hat{p},\, p,\,\bar{p}\right)$
to describe covariant vectors in light-cone coordinates, which
are related to Minkowski coordinates by

\begin{equation}
\check{p}=\left(p_{0}-p_{3}\right),\,\hat{p}=\left(p_{0}+p_{3}\right),\, p=\left(p_{1}-ip_{2}\right),\,\bar{p}=\left(p_{1}+ip_{2}\right).\end{equation}

In light-cone coordinates the metric becomes off-diagonal. The
Lorentz invariant product of two vectors is given by

\begin{equation}
p\cdot q=\left(\check{p}\hat{q}+\hat{p}\check{q}-p\bar{q}-\bar{p}q\right)/2.\end{equation}

To keep the derivation simple, the momentum components $p_{n\,\mu}$
of the $n^{th}$ external leg are simply denoted by number $n$ with
the appropriate decoration $\left(\check{n},\,\hat{n},\,\tilde{n},\,\bar{n}\right)$.
Note that a tilde is used for the $p=p_{1}-ip_{2}$ component to avoid
possible confusion with numerical factors.

A 4-vector can be written in the form of a bispinor 

\begin{equation}
P_{\alpha\dot{\alpha}}=p^{\mu}\sigma_{\mu\alpha\dot{\alpha}}=\left(\begin{array}{cc}
\check{p} & -p\\
-\bar{p} & \hat{p}\end{array}\right)\end{equation}

by contracting with $\sigma_{\mu}=\left(I_{2},\,\vec{\sigma}\right)$,
where $I_{2}$ is the $2\times2$ identity matrix and $\vec{\sigma}$
stands for Pauli matrices.

If $p_{\mu}$ is lightlike, $\check{p}=p\bar{p}/\hat{p}$ and the
bispinor factorises $p_{\alpha\dot{\alpha}}=\lambda_{\alpha}\bar{\lambda}_{\dot{\alpha}}$,
where

\begin{equation}
\lambda_{\alpha}=\left(\begin{array}{c}
-p/\sqrt{\hat{p}}\\
\sqrt{\hat{p}}\end{array}\right),\,\bar{\lambda}_{\dot{\alpha}}=\left(\begin{array}{c}
-\bar{p}/\sqrt{\hat{p}}\\
\sqrt{\hat{p}}\end{array}\right).
\label{spinors}\end{equation}

Spinors $\lambda_{i\alpha}$ associated with different massless particles
can be contracted to given a Lorentz invariant angle bracket

\begin{equation}
\left\langle 12\right\rangle =\epsilon^{\alpha\beta}\lambda_{1\alpha}\lambda_{2\beta}=\frac{\left(12\right)}{\sqrt{\hat{1}\hat{2}}},
\label{spinor product}\end{equation}

and we define a round bracket as 

\begin{equation}
\left(12\right)=\hat{1}\tilde{2}-\hat{2}\tilde{1}.\end{equation}


\end{document}